\documentclass[12pt]{article}
\usepackage{amsmath,amsfonts,bm}

\renewcommand{\baselinestretch}{1.4}
\newcommand{\CP}[1]{\mathbb{P}^{#1}}
\newcommand{\C}[1]{\mathbb{C}^{#1}}

\def\Z{\mathbb{Z}}

\def\be{\begin{equation}}
\def\ee{\end{equation}}
\def\bear{\begin{eqnarray}}
\def\eear{\end{eqnarray}}
\def\bV{\mathbf{V}}

\def\bX{\mathbf{X}}
\def\Y{\mathbf{Y}}
\def\bS{\mathbf{S}}

\def\dim{\mathrm{dim}}
\def\Vol{\mathrm{Vol}}

\def\Tr{\mathrm{Tr \,}}

\def\cale{\mathcal{E}}
\def\calf{\mathcal{F}}
\def\calh{\mathcal{H}}
\def\ch{\mathrm{ch}}
\def\Td{\mathrm{Td}}
\def\bi{\bibitem}
\def\Ext{\mathrm{Ext}}
\def\Hom{\mathrm{Hom}}
\usepackage{graphicx}

\begin{document}

\begin{titlepage}

\begin{flushright}
hep-th/0310262\\
NSF-KITP-03-89
\end{flushright}
\vfil

\begin{center}
{\huge Exceptional Collections and}\\
\vspace{3mm}
{\huge del Pezzo Gauge Theories} \\
\end{center}

\vfil
\begin{center}
{\large Christopher P. Herzog}\\
\vspace{1mm}
Kavli Institute for Theoretical Physics,\\
University of California, Santa Barbara, CA  93106, USA\\
{\tt herzog@kitp.ucsb.edu}\\
\vspace{3mm}
\end{center}

\vfil

\begin{center}

{\large Abstract}
\end{center}

\noindent
Stacks of D3-branes placed at the tip of
a cone over a del Pezzo surface provide a way
of geometrically engineering a small but rich class of
gauge/gravity dualities.
We develop tools for understanding the resulting
quiver gauge theories
using exceptional collections.  
We prove two important results for a general quiver
gauge theory: 1) we show the ordering of the nodes
can be determined up to cyclic permutation and 2)
we derive a simple formula for the ranks of the 
gauge groups (at the conformal point) 
in terms of the numbers
of bifundamentals.
We also provide a detailed analysis of four node
quivers, examining when precisely mutations
of the exceptional collection are related to Seiberg duality.
\vfil
\begin{flushleft}
October 2003
\end{flushleft}
\vfil
\end{titlepage}
\newpage
\renewcommand{\baselinestretch}{1.1}  

\renewcommand{\arraystretch}{1.5}

\section{Introduction}

A convenient way of engineering ${\mathcal N}=1$ gauge/gravity
dualities starts with a stack of D3-branes placed at the tip
of a Calabi-Yau cone $\bX$ \cite{MoPle,acharya,KW,Kehag}.
This construction generalizes the original AdS/CFT 
correspondence \cite{jthroat,gkp,EW}, where the D3-branes are
placed in flat space, $\bX = \C{3}$.  The resulting collection of 
gauge/gravity dualities is extremely rich; the number of 
qualifying $\bX$ is infinite.  Unfortunately, we lack a detailed
understanding of most of these $\bX$.  For example,
to the author's knowledge, the metric for only two such $\bX$
is known, $\C{3}$ and the conifold.  In this paper, we will
study dualities where $\bX$ is a (complex) 
cone over a del Pezzo surface.

A del Pezzo surface is a two complex dimensional, Kaehler manifold
with positive curvature.  Two simple examples of del Pezzos are
$\CP{2}$ and $\CP{1} \times \CP{1}$, for which we do know metrics.
The corresponding $\bX$ are orbifolds of $\C{3}$ and the conifold.
The remaining del Pezzos, denoted $dP_n$, 
correspond to $\CP{2}$ blown up at $n$
points where $1 \leq n \leq 8$.  

Even though 
metrics for the $dP_n$ are lacking, these surfaces
are extremely well studied and we can make progress in
understanding their gauge/gravity dualities.
For example, for $n=1,2,3$, the resulting $dP_n$ are toric
and the
gauge theories for these models can be extracted
using toric geometry \cite{beasley0,Beasley1,fhh1, fhh2}.

Exceptional collections appear to be one of the most promising
tools for understanding del Pezzo gauge theories.
Exceptional collections exist for all del Pezzos.  Given an
exceptional collection, it is easy to generate another
such collection through a braiding operation called 
mutation.  For every exceptional
collection, there is a prescription for writing
down the quiver, the ranks of the gauge groups, and
the R-charges of the fields at the conformal point.
A brief review of these collections is provided in section 2.

Exceptional collections were first 
proposed in this context by \cite{unify}
although the authors provided detailed analysis only
for toric cases.  
Later Wijnholt \cite{wijn} developed this proposal, 
deriving gauge theories for $n>3$.\footnote{
Other techniques have been proposed for deriving
the gauge theory for $n>3$: ``unhiggsing'' was suggested
by \cite{ffhh} and $pq$-seven brane webs by \cite{haiq}.}
Most recently, the author and Walcher \cite{HW} used
exceptional collections to understand the dibaryon spectra in
del Pezzo gauge theories, extending earlier work in \cite{HM, IW2}.    

Nevertheless, our understanding of the connection
between these collections and the del Pezzo gauge
theories is far from complete.  For example, it does
not appear to be true that every exceptional collection
generates a reasonable gauge theory.  Certain
collections generate gauge theories with gauge
groups of zero rank and bifundamentals with 
negative R-charge!  In this
paper, we try to fill some of the gaps in our
understanding.
 
One open problem is, given a quiver gauge theory
for a del Pezzo, can we reverse engineer the 
corresponding exceptional collection?  The
exceptional collection is an ordered collection 
of sheaves, and each sheaf is identified with
a node in the quiver.  Thus to generate
the exceptional collection, we have to 
order the nodes of the quiver.  

In section 3,
we partially solve this ordering problem. 
An exceptional collection, through a particularly
simple set of mutations, can generate a bi-infinite
sequence of sheaves called a helix.
We prove that every quiver corresponds to a helix
and that a cyclic permutation of the nodes
corresponds to choosing a different foundation for
the helix.  This result reduces the full ordering
problem to ordering the quiver up to cyclic permutation.

In section 4, we present a formula for the
gauge group ranks that
partially solves another part of the reverse
engineering problem.  
In \cite{HW}, these ranks 
were identified
with the ranks of the bundles in the ``dual''
exceptional collection.  
Our formula depends only on the numbers
of bifundamentals and an ordering of the quiver
and thus can be used in constructing the
``dual" collection.

In this paper, we
rechristen the ``dual'' collection the
geometric collection.  The original exceptional
collection we call the gauge theory collection.  

Finally, in section 5, we present a detailed 
analysis of four node quivers
for del Pezzos.  We will see
that there is only one such quiver, and only
two orderings (up to cyclic permutation)
are allowed.  We denote these orderings
$A$ and $F$ type.  Seiberg duality
of an $A$ type quiver produces another
$A$ type quiver.  Seiberg duality
of an $F$ type quiver either produces a
new $F$ type quiver or a quiver
that cannot be described by an exceptional
collection on a del Pezzo.
Inspired by these results, 
we define a 
{\it well split} quiver to be an ordered quiver
(corresponding to a helix)
where Seiberg duality on any node
is equivalent to a sequence of mutations.
Our $A$ type quiver is
well split.
More generally, an unproven conjecture is that
the Seiberg dual of a well split quiver
is again well split.  
In contrast, the $F$ type quiver is
{\it ill split} or equivalently not well split.
Leaving precise definitions for
the text, we believe it to be true that 
Seiberg
duality 
cannot be expressed in terms of mutations
for ill split quivers.

Ultimately, we hope that the current work
can be used to understand generalizations
of the Klebanov-Strassler (KS) solution \cite{klst}
where in addition to D3-branes, we add D5-branes
wrapped on vanishing two-cycles in $\bX$.  These
D5-branes take the theory away from the conformal point,
and the gauge couplings begin to run.  A qualitative
understanding of this RG flow can be gained by
performing a Seiberg duality every time a coupling
diverges.  For example, one can learn something
about the change in the number of degrees of 
freedom as a function of RG scale.
In this context, the analysis will be simpler if we
restrict to well split quivers.  We hope to 
return to these generalized KS flows in a future
publication \cite{FHHW}.

\section{Quivers from Exceptional Collections}

We review the construction of an ${\mathcal N}=1$ quiver gauge theory
from an exceptional collection, as described in \cite{wijn, HW}.
We will be interested in the class of gauge theories that are dual, via
the AdS/CFT correspondence, to type IIB string theory in an
$AdS_5 \times \Y$ background, where $\Y$ is a U(1) bundle over
a del Pezzo surface.

The starting point is an exceptional collection of sheaves $\cale$ 
on a del Pezzo $dP_m$, $m=0, \ldots, 8$.  
A standard mathematical reference for these collections is
\cite{rudakov} (see also \cite{NK}).
For the string theorist \cite{unify, hiv},
these sheaves are simply a set of elementary ``rigid'' branes generating
all BPS configurations of the theory by bound state formation.  
There exist special maps between the sheaves denoted $\Ext^i$, which
the string theorist may think of as the ground states of the strings 
connecting the elementary branes.  For each sheaf, we have a 
$SU(N)$ gauge group where $N$ corresponds to how many
of that particular type of brane we decided to include in the geometry.
Moreover, for each $\Ext^i$ map, we have bifundamental matter fields.

\subsection{A Review of Exceptional Collections}

Having given the rough picture, we now become precise.  Let $\bV$ be
a complex Fano variety, e.g. a del Pezzo.  A sheaf $E$ over $\bV$ is
called exceptional if $\Ext^0(E,E) = \Hom(E,E)=\C{}$ and $\Ext^k(E,E)=0$ for
$k>0$.  An ordered collection of sheaves $\cale = (E_1, E_2, \ldots, E_n)$
is called exceptional if each $E_i$ is exceptional and if,
moreover, for each pair $E_i$, $E_j$ with $i>j$, we have
$\Ext^k(E_i, E_j) =0$ for all $k$ and $\Ext^k(E_j, E_i) = 0$ except
possibly for a single $k$.  

To count the number of bifundamental fields in the gauge theory,
we must understand these $\Ext$ maps.  A useful tool is the generalized
Euler character
\be
\chi(E,F) = \sum_i (-1)^i \dim \Ext^i(E,F)
\ee
which by the Hirzebruch-Riemann-Roch theorem can be rewritten as
\be
\chi(E,F) = \int_\bV \ch(E^*) \ch(F) \Td(\bV)
\label{chiR}
\ee
where $\ch(E)$ is the Chern character of the sheaf $E$.
Note that $\chi$ is a bilinear form.  
An additional useful fact for del Pezzos
is that only $\Ext^1$ and $\Ext^0$ can be nontrivial \cite{KO}.  
This fact together with the Euler character allows one to compute
the numbers of bifundamentals in the gauge theory quickly and
easily from the Chern characters of the exceptional sheaves.

An obvious question at this point is how many sheaves do we include
in the exceptional collection, or equivalently how many 
fundamental branes do we
need to describe the physics.  Geometrically, in these del Pezzos, 
branes can correspond to points, they can wrap curves, or they
can wrap the entire del Pezzo.  Thus, the number $n$ of sheaves
in the collection
should correspond to the sum of the Betti numbers of $dP_m$,
$n=m+3$.
Mathematically, we see that the Chern
character $\ch(E) = (r(E), c_1(E), \ch_2(E))$ is described by $n$ charges.
At the level of Chern characters, we can have at most $n$ linearly
independent sheaves in our collection.  A complete exceptional
collection contains $n$ sheaves and spans this $n$-dimensional
vector space.

In components, the Euler character reads
\begin{eqnarray}
\chi(E,F) &=& r(E)r(F) + \frac{1}{2} (r(E) \deg(F) - r(F) \deg(E)) \nonumber \\
&& + r(E) \ch_2(F) + r(F) \ch_2(E) - c_1(E) \cdot c_1(F) \ , 
\label{chiB}
\end{eqnarray}
which can easily be derived from (\ref{chiR}) using
$\Td(dP_n) = 1 - \frac{K}{2} + H^2$, where $K$ is the canonical
class and $H$ is the hyperplane, with $\int_{dP_n} H^2 = 1$.
Also the degree $\deg(E) = (-K) \cdot c_1(E)$.

If $\cale$ is an exceptional collection, one obtains new 
exceptional collections (and hence new gauge theories)
by left and right mutations:
\begin{eqnarray}
L_i : (\ldots, E_{i-1}, E_i, E_{i+1}, \ldots ) &\to& 
(\ldots, E_{i-1}, L_{E_i} E_{i+1}, E_i, \ldots)  \ ,\nonumber \\
R_i: (\ldots, E_{i-1}, E_i, E_{i+1}, \ldots ) &\to& 
(\ldots, E_{i-1}, E_{i+1}, R_{E_{i+1}} E_i, \ldots) \ .
\end{eqnarray}
Here, $L_{E_i} E_{i+1}$ and $R_{E_{i+1}} E_i$ are defined by
short exact sequences, whose precise form depends on which
of the $\Ext^k(E_i, E_{i+1})$ are non-zero.  At the level of
the Chern character
\begin{eqnarray}
\ch(L_E F) &=& \pm (\ch(F) - \chi(E,F) \ch(E)) \ , \nonumber \\
\ch(R_F E) &=& \pm(\ch(E) - \chi(E,F) \ch(F)) \ ,
\label{mutate}
\end{eqnarray}
where the sign is chosen such that the rank of the mutated
bundle is positive.  We introduce some additional nomenclature
here that will be important later on.  If $\chi(E,F)<0$, the
mutation is called an {\it extension}
and the plus sign above is chosen.  
In the remaining cases, choosing the plus sign corresponds
to a {\it recoil} while the negative sign is a {\it division}.

There are an additional class of mutations denoted $L^D$ and
$R^D$, which at the level of charges leads to the selection of
the plus sign in (\ref{mutate}) above.  Choosing the plus
sign will lead often to sheaves with negative rank, which
roughly speaking one may think of as the antibrane.
For more details on mutations, see for example \cite{rudakov}.

\subsection{Constructing the Gauge Theory}

We review the construction of a 
${\mathcal N}=1$ superconformal
gauge theory dual to string theory on $AdS_5 \times \Y$ for
$\Y$ a U(1) bundle over $dP_m$.

To construct the gauge theory, we begin with an exceptional
collection $\cale^G = (E^G_1, E^G_2,\ldots,  E^G_n)$ over $dP_m$.
The quiver will consist of $n$ nodes, one node
for each $SU(N_i)$ gauge group.  
The ranks of the gauge groups are defined to be $N_i = r(E^G_i)N$.
Such a collection $\cale^G$ we will refer to as a geometric collection.

Next we construct the dual exceptional collection $(\cale^Q)^\vee = \cale^G$.
For any exceptional collection $\cale$, we define
the dual collection $\cale^\vee$ to be the result of a
braiding operation,
\begin{eqnarray}
\cale^\vee &= & (E_n^\vee, E_{n-1}^\vee, \ldots, E_1) \nonumber \\
&=& (L_{E_1}^D \cdots L_{E_{n-1}}^D E_n, L_{E_1}^D \cdots L_{E_{n-2}}^D E_{n-1},
\ldots, L_{E_1}^D E_2, E_1) \ .
\end{eqnarray}
 The collection $\cale^\vee$ is exceptional in the order presented, and
is dual to $\cale$ in the sense of the Euler form, i.e. 
$\chi(E_i, E_j^\vee) = \delta_{ij}$.  Note that because of the
D-type mutations involved, the ranks of $\cale^\vee$ may not
be all positive.  We call this dual collection $\cale^Q$ the
gauge theory collection.  Note that the superscript 
$\vee$ is
meant only to indicate the dual.  

Third, we construct the incidence matrix
\be
S_{ij} = \chi(E^Q_j, E^Q_i) \ .
\ee
From the exceptional property, $S$ will be an upper triangular
matrix with integer entries and ones along the diagonal.
The incidence matrix for the original geometric collection is 
the inverse of $S$.  Note that $S^{-1}$ is also upper triangular,
integer valued, and has ones along the diagonal.

We use $S$ to compute the numbers of bifundamentals.
We draw $S_{ij}$ arrows from node $i$ to node $j$ in the quiver.
(A negative $S_{ij}$ means we have to reverse the direction of 
the arrows.)  Each arrow is a bifundamental ${\mathcal N}=1$
chiral superfield transforming under the gauge groups
at the tail and head of the arrow.  It is straightforward to verify
that the chiral anomalies in the resulting gauge theory vanish:
\be
\sum_j (S-S^T)_{ij}  r(E^G_j) = 0 \ .
\label{sst}
\ee

Finally, we may compute the R-charges of the bifundamental fields.
Let $i$ correspond to the tail of the arrow and $j$ to the head.
Building on work of Intriligator and Wecht \cite{IW2}, Herzog 
and Walcher \cite{HW} showed that 
\be
R(X_{ij}) = \frac{2}{K^2 r(E^G_i) r(E^G_j)} \times
\left\{ 
\begin{array}{cc}
\chi(E^G_i, E^G_j) & \mathrm{if } \; \; \chi(E^G_i, E^G_j) \neq 0 \\
\chi(E^G_i \otimes K, E^G_j) & \mathrm{otherwise.}
\end{array}
\right.
\label{Req}
\ee
From these R-charges, one may verify that the NSVZ beta functions vanish.
In particular, it was shown in \cite{HW} that
\be
0 = \beta_i =  \frac{1}{K^2} \sum_k (\chi(E_i^Q, E_k^Q) - \chi(E_k^Q, E_i^Q))
(\chi(E_i^G, E_k^G) - \chi(E_k^G,E_i^G)) + r(E_i^Q) r(E_i^G) \ .
\label{beta}
\ee

We can move away from the conformal point by adding fractional
branes to the geometry.  The fractional branes 
change the ranks of the gauge groups in a way that continues
to preserve the cancellation of the chiral anomalies
(\ref{sst}).  Thus,
these fractional branes correspond
to the remaining vectors in the kernel of $(S-S^T)$.

\section{Helices and Quivers}

Having constructed a conformal gauge theory starting from 
an exceptional collection,
it is natural to wonder whether an exceptional collection can be constructed
from a conformal gauge theory.   Figuring out the ordering of the nodes is clearly
important.  The quiver does not suggest any obvious ordering.

As a first step to solving the ordering problem, we will show that the choice of
ordering should be independent of cyclic permutations of all the nodes.  For
example, for a four node quiver, the ordering $(1234)$ should be equivalent
to the ordering $(2341)$.
The proof requires introducing the notion of a helix.

A helix $\calh = (E_i)_{i\in \Z}$ is a bi-infinite
extension of an exceptional collection $\cale$ 
defined recursively by
\begin{eqnarray}
E_{i+n} &=& R_{E_{i+n-1}} \cdots R_{E_{i+1}} E_i \ , \nonumber \\
E_{-i}  &=& L_{E_{i-1}} \cdots L_{E_{n-1-i}} E_{n-i} \qquad i\geq 0 \ ,
\end{eqnarray}
such
that the helix has period $n$, by which we mean
\be
E_i = E_{n+i} \otimes K \qquad  \forall i \in \Z \ .
\ee
A complete exceptional collection on a del Pezzo
will generate a helix.
Moreover, the helix remains the same 
(for our purposes) if constructed
instead with $D$-type mutations.
Any subcollection of $\calh$ of the form
$(E_{i+1}, E_{i+2}, \ldots, E_{i+n})$ is exceptional and is
called a foundation of $\calh$.

Given a geometric 
collection $\cale^G$ and the associated helix $\calh$,
a natural question is how does the gauge theory
depend on the choice of foundation.  The answer is
that it doesn't.  The gauge theory depends only on the choice of
helix.  
In other words, for every helix, there exists a unique quiver.
Moreover, shifting the foundation corresponds on 
the gauge theory side to a cyclic permutation of the nodes of
the quiver.

Let $\cale = (E_1, E_2, \ldots, E_n)$ and $\calf = (E_n \otimes K, E_1, E_2, \ldots, E_{n-1})$
be two neighboring geometric foundations of $\calh$.
Tensoring with $K$ does not affect the rank of the sheaf.
Thus, the ranks of the gauge groups will be cyclically permuted but otherwise unchanged.
We can prove that the quiver is independent of the helix by showing that
the quivers constructed from 
the gauge theory collections $\cale^\vee$ and $\calf^\vee$ are 
identical.  

Consider the dual exceptional collections
$\cale^\vee = (E_n^\vee, E_{n-1}^\vee, \ldots, E_1^\vee)$ and
$\calf^\vee$:
\be
\cale^\vee = (E_n \otimes K, L^D_{E_1} L^D_{E_2} \cdots L^D_{E_{n-2}} E_{n-1}, \ldots,
L^D_{E_1} E_2, E_1)
\ee
and
\be
\calf^\vee = (E_{n-1} \otimes K, 
L^D_{E_n \otimes K} L^D_{E_1} L^D_{E_2} \cdots L^D_{E_{n-3}} E_{n-2}, \ldots,
L^D_{E_n \otimes K} E_1, E_n \otimes K) \ .
\ee

To analyze these dual collections, we need a couple of lemmas.
Let $G$, $E$, and $F$
be three exceptional sheaves.
It follows from linearity of the Euler character $\chi$ and the definition of
left mutation that 
\be
\chi(L^D_G E, F) = \chi(E - \chi(G,E) G, F) = \chi(E,F) - \chi(G,E)\chi(G,F) \ .
\ee
In the case $G=F$ and if $\chi(E,G)=0$, it follows that 
\be
\chi(L^D_G E, G) = - \chi(G,E) \ . 
\label{quiveridt}
\ee
If $\chi(E,G)=0$, $\chi(F,G)=0$,
and $\chi(F,E)=0$, it follows that
\be
\chi(L^D_G E, L^D_G F) = \chi(E,F) \ .
\label{quiverido}
\ee

Let $S_{ij}=\chi(E^\vee_{n+1-i}, E^\vee_{n+1-j})$ 
and $T_{ij} = \chi(F^\vee_{n+1-i}, F^\vee_{n+1-j})$
be $n \times n$ matrices constructed from $\cale^\vee$ 
and $\calf^\vee$ respectively.
Consider the submatrix $\chi(E^\vee_{n+1-i}, E^\vee_{n+1-j}) = s_{ij}$ where 
$E^\vee_i$ and $E^\vee_j$ can be any sheaves 
in $\cale^\vee$ except for $E_n \otimes K$.
Let $t_{ij}$ be the corresponding submatrix  for $\calf^\vee$
where again we are not allowed to use $E_n \otimes K$.  
The submatrices $s$ and $t$
are identical as $(n-1) \times (n-1)$ dimensional matrices.  This
statement follows from (\ref{quiverido}).

Now consider the remaining entries in the $T$ and $S$ matrices.  In
particular, consider $S_{1j}$ and $T_{(j-1)n}$ where $j=2,3,\ldots,n$.
($S_{11}=T_{nn}=1$ and the other entries vanish trivially because
of the ordering inside the collection.)
It follows from (\ref{quiveridt}) that
\begin{eqnarray}
T_{(j-1)n} &=& 
\chi(L^D_{E_n \otimes K} L^D_{E_1} \cdots L^D_{E_{n-j}} E_{n-j+1}, E_n \otimes K) 
\nonumber \\
&=&
-\chi(E_n \otimes K, L^D_{E_1} \cdots L^D_{E_{n-j}} E_{n-j+1})
\nonumber \\
&=& -S_{1j} \ .
\label{lrel}
\end{eqnarray}

Using the matrices $T$ and $S$ we can construct quivers.  
These quivers will be identical up to a cyclic permutation 
of the
nodes.  Clearly the quivers from the submatrices $s$ and $t$ must
be identical.  The minus sign in (\ref{lrel}) then compensates
for the cyclic permutation.  

\subsection{Ordering and the Superpotential}

These del Pezzo gauge theories can have a superpotential which up to
this point we have ignored.  The superpotential, if known, 
further constrains the ordering of the nodes.

The superpotential $W$ is a gauge invariant polynomial in the bifundamentals
$X_{ij}$ of the quiver gauge theory.  The superpotential generates relations
in the path algebra of the quiver:
\be
\frac{\partial W}{\partial X_{ij}} = 0 \ .
\ee
Because of the R-symmetry, $W$ will have R-charge two.  A typical term
in the superpotential corresponds to a loop in the quiver.  One multiplies
the associated $X_{ij}$ together and traces over the color indices.

From the R-charge formula (\ref{Req}), it should be clear that a loop in the
quiver will produce a monomial in the $X_{ij}$ with an R-charge
that is a positive integer multiple of two \cite{HW}.  
Assume for the moment that we know the correct ordering of the
quiver.  If we take the convention where
$i$ corresponds to the tail of the arrow and $j$ to the head, we get an
additional two in the R-charge every time $j>i$ in the monomial.  For example,
$X_{43}X_{32}X_{21}X_{14}$ would have R-charge two while
$X_{34}X_{42}X_{21}X_{13}$ would have R-charge four.  

Working backward, we see that the order the nodes appear in monomials in the superpotential
must be the same order in which the nodes appear in the exceptional collection.
The superpotential is often enough to specify the order of the nodes in 
the collection up to cyclic permutation.

\section{The Ranks of the Gauge Groups}

Having partially solved the ordering portion of the
inverse problem, we now
derive a formula for the ranks of the gauge groups (or
equivalently the ranks of the sheaves in the helix) from
the numbers of bifundamentals.  This formula can
be used to constrain further the ordering 
as we will see in section 5.

We assume the existence of a gauge theory
collection $\cale^Q = (E_1, \ldots, E_n)$.
From the formula for the Euler character (\ref{chiB}),
we know that
\be
\chi(E_1^\vee \otimes K, E_j^\vee) = K^2 r(E_1^\vee) r(E_j^\vee)
- \chi(E_j^\vee, E_1^\vee) \ .
\label{firstr}
\ee
Note that the Euler character is invariant under tensoring both sheaves 
with an invertible sheaf.  Some standard manipulation gives
\begin{eqnarray*}
\chi(E_1^\vee \otimes K, E_j^\vee) &=& 
\chi(E_1^\vee, E_j^\vee \otimes (-K)) \\
&=& \chi(E_1, (L^D_{E_1} \cdots L^D_{E_{j-1}} E_j) \otimes (-K)) \\
&=& \chi(E_1, R^D_{E_n} \cdots R^D_{E_{j+1}} E_j) \ .
\end{eqnarray*}
Similarly for the other term in (\ref{firstr}), one finds
\begin{eqnarray*}
\chi(E_j^\vee, E_1^\vee) &=& \chi(L^D_{E_1} \cdots L^D_{E_{j-1}} E_j, E_1) \\
&=& \chi(L^D_{E_2} \cdots L^D_{E_{j-1}} E_j, E_1) - \chi(E_1, E_1) 
\chi(E_1, L^D_{E_2} \cdots L^D_{E_{j-1}} E_j) \\
&=& -\chi(E_1, L^D_{E_2} \cdots L^D_{E_{j-1}} E_j) \ . 
\end{eqnarray*}
Putting these two small results together, one finds for (\ref{firstr}) that
\be
K^2 r(E_1^\vee) r(E_j^\vee) =
\chi(E_1, R^D_{E_n} \cdots R^D_{E_{j+1}} E_j)
-\chi(E_1, L^D_{E_2} \cdots L^D_{E_{j-1}} E_j) \ .
\label{secr}
\ee
This formula (\ref{secr}) has a simple graphical interpretation.  We put $n$ points
on a circle and label them clockwise from one to $n$.  The second term in
(\ref{secr}) is a sum over paths from point one to point $j$, while the first term is a sum
over paths from point $j$ to point one.

More precisely, let ${\mathcal P}_{ab}$ be the set of paths from $a$ to $b$, with
$a<b$.
Let $I \in {\mathcal P}_{ab}$ be a particular path.  $I$ is a map from the set of integers
$0\leq j \leq m+1$ to the set of integers $a \leq k \leq b$, $I(j) = k_j$, such that
if $i<j$ then $k_i < k_j$.  Moreover, $k_0 = a$ and $k_{m+1} = b$.  Finally, 
the path length of $I$ is defined to be $C(I) = m$. 

With this notation, we may write
\be
K^2 r(E^\vee_i) r(E^\vee_j) = \sum_{I\in {\mathcal P}_{i,j} \cup {\mathcal P}_{j,i+n}} 
(-1)^{C(I)+1} \prod_{s=0}^{C(I)} x_{I(s),I(s+1)} 
\label{rankpath}
\ee
where we have defined $x_{kl} = (S-S^T)_{kl}$ with $S_{kl} = \chi(E_k, E_l)$.
We put the whole set of paths on a circle by
identifying $x_{l,k+n} = -x_{kl}$. 
We have used the cyclic permutativity proved in the previous section to
substitute an arbitrary node $i$ for $1$ in (\ref{rankpath}).  Indeed, the 
formula remains valid even when $i=j$, as we will see below.

To gain confidence that we have the right formula, we check that the
chiral anomalies cancel
\be
K^2 r_i \sum_{j=1}^n r_j x_{ij} = 0 \ ,
\label{chan}
\ee
where $r_i = r(E_i^\vee)$ are the ranks of the gauge groups.
The $x_{ij}$ closes all the paths from $i$ to $j$ to make loops.
We will see that this sum over loops vanishes in a trivial way.  
By a loop we mean an element $I \in {\mathcal P}_{j,j+n}$ together
with the obvious equivalence relation.  In particular,
two paths $I \in {\mathcal P}_{j,j+n}$ and $I' \in {\mathcal P}_{j', j'+n}$
construct the same loop if they have the same length $C(I) = C(I')$ and when
viewed on the circle, visit the same nodes in the same cyclic order.

Let ${\mathcal L}_i$ be the set of loops which include the node $i$.
This set of loops can be decomposed into a sum over paths.  In particular
\be
{\mathcal L}_i = \bigoplus_{k=1}^{n-1} 
\overline{{\mathcal P}_{i,i+k}} \ .
\ee
The overline notation means that we have closed the path into a loop by joining
nodes $i$ and $i+k$. More precisely, let $I \in {\mathcal P}_{ab}$.
Then ${\overline I} \in {\mathcal P}_{a,a+n}$.  For $0\leq j \leq m+1$,
$I(j) = {\overline I}(j)$ and ${\overline I}(m+2)=a+n$.
We can also perform this decomposition of loops into paths in a different way:
\be
{\mathcal L}_i = \bigoplus_{k=1}^{n-1} 
\overline{{\mathcal P}_{k+i,i+n}} \ .
\ee

Now we substitute (\ref{rankpath}) into (\ref{chan}) and switch the order
of summation.  We find that we have summed over all loops involving
the node $i$ twice.  The $x_{ij}$ in (\ref{chan}) introduces a relative
minus sign between the two sums.  Each loop in one sum pairs up with
a loop in the other and cancels.

As a final check, we derive a formula for $K^2 r_1^2$ using the NSVZ beta function $\beta_1$.  
As reviewed above, it was demonstrated in \cite{HW} that 
\be
0 = \beta_1 = r(E_1^\vee)^2 + \frac{1}{K^2} \sum_k 
(\chi(E_1, E_k) - \chi(E_k, E_1)) 
(\chi(E_1^\vee, E_k^\vee - \chi(E_k^\vee, E_1^\vee)) \ ,
\label{beta1}
\ee
where we have used the fact that $E_1^\vee = E_1$.
This proof is more rigorous than it may seem.  We are not assuming
that the beta function vanishes in order to derive $r_1$.  Rather,
in \cite{HW}, the right hand side of (\ref{beta1}) was shown
to vanish as a mathematical identity.

From the beta function then
\begin{eqnarray}
K^2 r_1^2 &=& \sum_k \chi(E_1, E_k) \chi(E_k^\vee, E_1^\vee) \nonumber \\
&=& -\sum_k \chi(E_1, E_k) \chi(E_1, L^D_{E_2} \cdots L^D_{E_{k-1}} E_k) \ .
\end{eqnarray}
Using the path language, the sum is over all loops involving the node 1.
Using cyclic invariance, we conclude that $K^2 r_i^2$ is in general
a sum over all loops involving the node $i$:
\be
K^2 r_i^2 = \sum_{L \in \mathcal{L}_i} (-1)^{C(L)+1} 
\prod_{s=0}^{C(L)} x_{L(s),L(s+1)} \ .
\ee  
Note that loops with even numbers
of $x_{kj}$  
appear with a plus sign.
This formula is equivalent to (\ref{rankpath}) in the case $i=j$.

There is a simpler way of writing this sum.  
Formally, we may change the basis of the $S$ matrix.
\[
S \rightarrow B S B^T \ .
\]
Although $S$ is not invariant under such a change
of basis, $\Tr S^{-1} S^T$ is, and we can use this 
trace invariant.
First we evaluate this trace invariant in a particularly simple
basis.
In particular, we choose the basis in which the sheaves
are written in terms of their rank, first chern class,
and second chern character, as in (\ref{chiB}).
It is straightforward to see that $\Tr S S^{-T} = n$.
Evaluating $\Tr S S^{-T}$ in the exceptional collection 
basis, we find a sum over all loops in the quiver.
In particular, the first row of $S$ times the first column
of $S^{-T}$ is a sum over all loops involving node one
plus a one from the diagonal elements.
Then the second row of $S$ times the second column
of $S^{-T}$ will be a sum over all loops involving
node two but not involving node one again plus
a one from the diagonal elements, and so on.
Comparing $\Tr S S^{-T}$ in the two bases, we find
that the sum over all loops must
vanish,
\be
0 = \sum_{L \in \mathcal{L}} (-1)^{C(L)} 
\prod_{s=0}^{C(L)} x_{L(s),L(s+1)} \ .
\ee  
Thus $K^2 r_i^2$ may also be thought of
as a sum over all loops not involving node $i$.
(Now loops with even numbers of $x_{kj}$ will appear with 
a minus sign.)

As a coda to this section, we rewrite our formula for
$K^2 r_i^2$ in yet one more way.
 In particular, we produce the minor $S^i$
of $S$ by crossing out the $i$th row and $i$th column.  
For a quiver
with $n+1$ nodes ($n < 12$), 
\be
K^2 r_i^2 = -\Tr (S^i)^{-1} (S^i)^T +n \ .
\label{rankminor}
\ee  

\section{The Four Node Quiver}

We closely investigate the inverse problem for four node quivers.
The starting point is a four node quiver with arrows joining all the nodes.  In order to satisfy anomaly cancellation, at each node there must be either two arrows in and one arrow out or two arrows out and one arrow in (assuming no arrows vanish).  

\begin{figure}
\includegraphics[width=6.5in]{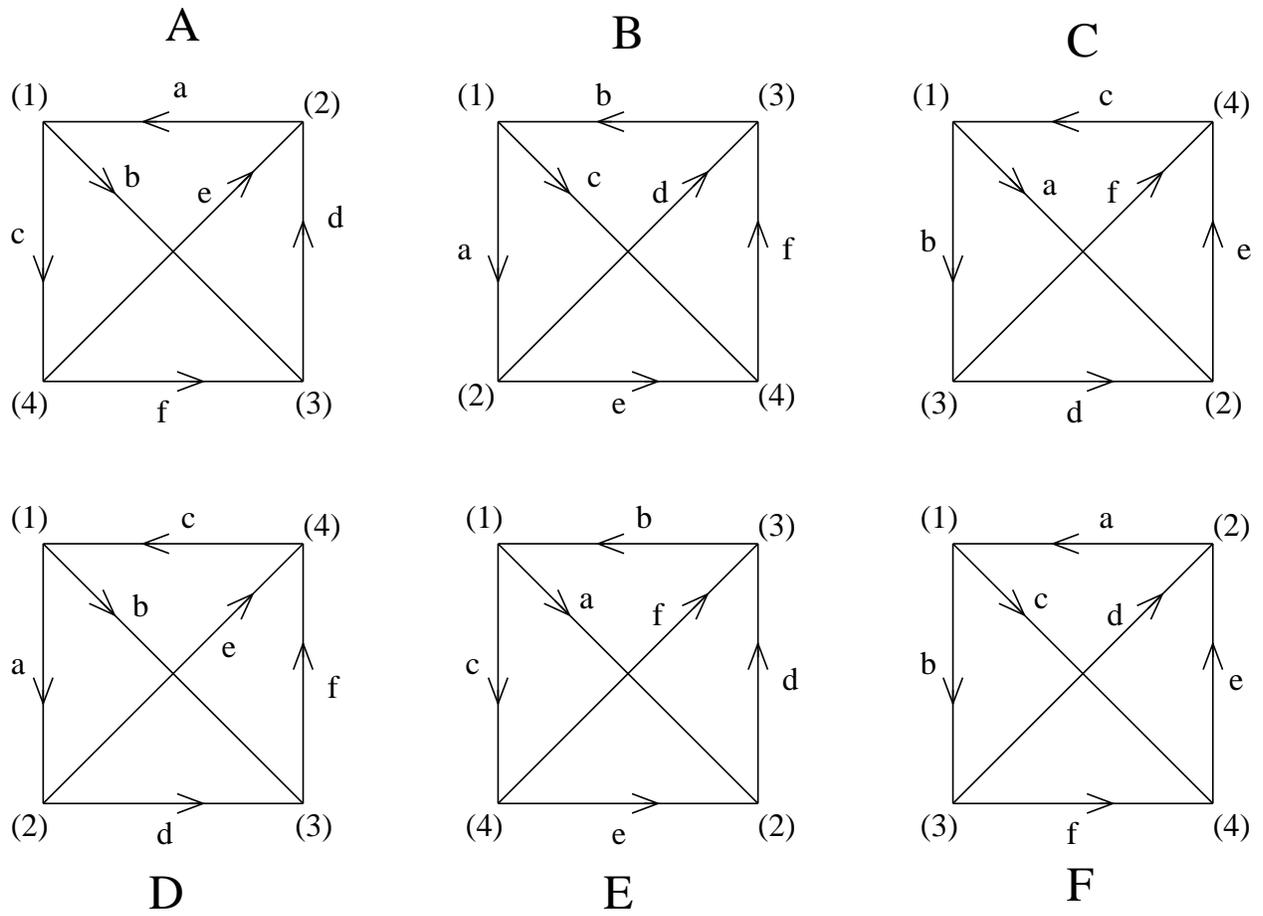}
\vfil
\caption{Different possible labelings of the four node quiver.}
\label{figfn}
\end{figure}

Our first lemma is that there is only one such quiver up to permutation of the nodes.  
We will draw this quiver as in figure \ref{figfn}.  
A brute force proof 
(which the writer employed) is to draw all the 
possibilities and relate them by permutation.

We will assume that this quiver corresponds to a geometric collection
$\cale^G = (E_4^\vee, E_3^\vee, E_2^\vee, E_1^\vee)$
with the ranks of the $E_i^\vee$ all positive and the numbers of
bifundamentals determined from the dual gauge theory
collection $\cale^Q$.
The nodes in the quiver, in some yet to be
determined order, correspond 
to the sheaves and the arrows to the $\Ext$ groups between the sheaves
in $\cale^Q$.  

A little bit of notation is in order.  We have the matrix 
$S_{ij} = \chi(E_i, E_j)$.  We write $S$ as
\be
S = \left( 
\begin{array}{cccc}
1 & a & b & c \\
0 & 1 & d & e \\
0 & 0 & 1 & f \\
0 & 0 & 0 & 1
\end{array}
\right)
\ee

From the previous section, it is clear that we need only care about the order up to cyclic permutations.  If the dual of 
$\cale^G = (E_4^\vee, E_3^\vee, E_2^\vee, E_1^\vee)$ generates our 
quiver, then so will the dual of 
$\calf^G = (E_1^\vee \otimes K, E_4^\vee, E_3^\vee, E_2^\vee)$. 

Without loss of generality, we may choose the node in the upper left hand
corner of figure \ref{figfn} to be node number one.  There are then six possible
orderings of the remaining nodes, corresponding to the six possible permutations
of three objects.
We label these six permutations $A$ through $F$.  
We will use a small Roman
numeral to denote the particular cyclic ordering.  For example $A$
type labeling corresponds to a clockwise labeling of the nodes.
Then $Ai$ corresponds
to the quiver with node one in the upper left and the nodes labeled 
clockwise, $Aii$ with node one in the 
upper right, $Aiii$ in the lower right, and $Aiv$ in the lower left.  
The $i$ type labelings always correspond 
to quivers with node one in the upper left and are shown in figure \ref{figfn}.

These orderings give us information about 
the signs of the entries of the $S$ matrix.  
We can ask if these sign assignments are consistent 
with what we expect for an exceptional collection.  
We will find that only the $A$ and $F$ type labelings are allowed.

The sign assignments for the labelings are as follows
\begin{eqnarray}
\begin{array}{|c|cccc|}
\hline
A & i & ii & iii & iv \\
\hline
a & - & - & - & - \\
b & + & + & - & - \\
c & + & + & + & + \\

d & - & - & - & - \\
e & - & + & + & - \\
f & - & - & - & - \\
\hline 
\end{array}
\qquad
&
\qquad
\begin{array}{|c|cccc|}
\hline
B & i & ii & iii & iv \\
\hline
a & + & - & - & + \\
b & - & - & + & + \\
c & + & + & - & - \\
d & + & + & - & - \\
e & + & - & - & + \\
f & - & + & + & - \\
\hline
\end{array}
\qquad
& 
\qquad
\begin{array}{|c|cccc|}
\hline
C & i & ii & iii & iv \\
\hline
a & + & + & + & - \\
b & + & - & - & + \\
c & - & - & + & - \\
d & - & + & + & + \\
e & + & + & - & - \\
f & + & - & + & + \\
\hline
\end{array}
\label{signassign}
\end{eqnarray}
\begin{eqnarray*}
\begin{array}{|c|cccc|}
\hline 
D & i & ii & iii & iv \\
\hline
a & + & + & + & + \\
b & + & - & - & + \\
c & - & - & - & - \\
d & + & + & + & + \\
e & + & + & - & - \\
f & + & + & + & + \\
\hline
\end{array}
\qquad
&
\qquad
\begin{array}{|c|cccc|}
\hline 
E & i & ii & iii & iv \\
\hline
a & + & - & - & + \\
b & - & + & + & - \\
c & + & + & - & - \\
d & + & + & - & - \\
e & - & - & + & + \\
f & - & + & + & - \\
\hline
\end{array}
\qquad
&
\qquad
\begin{array}{|c|cccc|}
\hline 
F & i & ii & iii & iv \\
\hline
a & - & - & + & - \\
b & + & + & - & - \\
c & + & - & + & + \\
d & - & - & - & + \\
e & - & + & + & - \\
f & + & - & - & - \\
\hline
\end{array}
\end{eqnarray*}

We begin by eliminating the $C$ type labeling.  
From the arguments presented above,
we know that
$\Tr S^{-1} S^T = 4$ is an invariant of
the exceptional collection.
Evaluating $\Tr S^{-1} S^T$ for arbitrary $S$, 
we find that
\be
a^2 + b^2 + c^2 + d^2 + e^2 + f^2 - abd - ace - bcf - def + acdf = 0 \ .
\label{I}
\ee
From the sign assignments in (\ref{signassign}), one can see that
every term in this trace invariant (\ref{I}) 
will be positive for the $C$ labeling and
that this trace invariant can never vanish.  Thus, we eliminate the $C$ labeling.

A little more information can be gleaned from our change of basis.
In the chern character basis, $S-S^T$ is clearly of rank two.  
The rank
will not change under change of basis.  Thus we conclude that
\be
cd - be + af = 0 \ .
\label{II}
\ee
Every sign assignment in (\ref{signassign}) is consistent with (\ref{II}) so
we will have to go farther afield to eliminate labelings
$B$, $D$, and $E$.

Our next step is to establish the ranks of the gauge groups, or equivalently
the ranks of the bundles in $\cale^G$.  
We can use the general formulae presented above.
For example, (\ref{rankminor}) tells us 
that
\begin{eqnarray}
8 r_1^2 & = & d^2 + e^2 + f^2 - def \ ,\nonumber \\
8 r_2^2 & = & b^2 + c^2 + f^2 - bcf \ , \nonumber \\
8 r_3^2 & = & a^2 + c^2 + e^2 - ace \ , \nonumber \\
8 r_4^2 & = & a^2 + b^2 + d^2 - abd \ .
\end{eqnarray}
In establishing the ordering, we found the
bilinears of the form $r_i r_j$, $i \neq j$
more useful.  The equation (\ref{rankpath})
gives
\begin{eqnarray}
8 r_1 r_2 & = & cdf - bd - ce\ ,  \nonumber \\
8 r_1 r_3 & = & ad - cf \ , \nonumber \\
8 r_1 r_4 & = & ae + bf - adf \ ,
\label{rels} \\
8 r_2 r_3 & = & acf - ab - ef \ , \nonumber \\
8 r_2 r_4 & = & -ac + fd \ , \nonumber \\
8 r_3 r_4 & = & acd - de - bc \ . \nonumber
\end{eqnarray}

The rank formulae (\ref{rels}) 
allow us to rule out orderings $B$, $D$, and $E$.  
Because the ranks of the gauge groups (and of the sheaves
in $\cale^G$) must be positive, the right hand side of the relations
(\ref{rels}) must be positive.  The sign assignments
of $B$, $D$, and $E$ would force some ranks to be negative.

We have succeeded in reducing the 24 different orderings of the nodes to two
possible orderings, the $A$ and $F$ type orderings.  Typically, to 
distinguish between the two orderings, it is enough to figure out
which ordering satisfies (\ref{I}).  

We can figure out the allowed terms in the superpotential by looking
at the R-charges of the bifundamentals.  
A term in the superpotential corresponds to a loop in the quiver where we
trace over the internal indices of the bifundamentals.  Looking at the four node
quiver, there are two triangular loops and one square loop that could
produce scalar superpotential terms.
The superpotential terms must have R-charge two.  
As discussed in section 3.1, the R-charge of a loop will be twice the number of
times neighboring nodes are not in descending order as we go around the loop.
For the $A$ type quiver, all three terms are allowed.  However, for the $F$ type
quiver, the square loop will have R-charge four and cannot appear in the superpotential.

\subsection{The Cubic R-charge Anomaly}

With these ranks, one can check the value of the 
cubic R-charge anomaly for the four node quivers.  
For ${\mathcal N}=1$ superconformal gauge theories,
the conformal anomaly \cite{anselmi1, anselmi2} is
\be
a_c = \frac{3}{32} (3 \Tr R^3  - \Tr R) \ .
\ee
The vanishing of the NSVZ beta functions tells us
that $\Tr R = 0$.  From AdS/CFT, we expect
\cite{gkp, skenderis, gubser} 
\be
a_c = \frac{\Vol(\bS^5)}{4 \Vol(\Y)} \ ,
\ee
where $\Y$ is either a $U(1)$ bundle over 
$\CP{1} \times \CP{1}$ or $dP_1$.
In both cases, we know that $\Vol(\Y) = 8\pi^3  /27$
\cite{BH}, and
so we expect that $\Tr R^3 = 3$.

The R-charges we know from (\ref{Req}):
\begin{eqnarray}
\Tr R^3 &=& \sum_{i=1}^4 r_i^2 + 
a(1-R_a)^3 r_1 r_2 + b(1-R_b)^3 r_1 r_3 
 + c(1-R_c)^3 r_1 r_4 \nonumber \\
&& + d(1-R_d)^3 r_2 r_3 + e(1-R_e)^3 r_2 r_4 + f(1-R_f)^3 r_3 r_4 
\end{eqnarray}
where
\be
R_a = \frac{-2a}{r_1 r_2} \ , \qquad R_d = \frac{-2d}{r_2 r_3} \ ,
\qquad R_f = \frac{-2f}{r_3 r_4} \ ,
\ee
\be
R_b = \frac{2 (ad-b)}{r_1 r_3} \ , \qquad
R_e = \frac{2 (df - e)}{r_2 r_4} \ ,
\ee
\be 
R_c = \frac{2 (-c + ae + bf - adf)}{r_1 r_4} \ .
\ee
This expression for $\Tr R^3$ is valid
for both $A$ and $F$ type quivers.
The expression is rather cumbersome, even for 
computer aided algebra, but with some care, we were
able to verify that $\Tr R^3 = 3$ when subject to 
the constraints (\ref{I}) and (\ref{II}).

Note that these ``R-charges'' $R_a$, $R_b$, and so on are not
the true R-charges.  Depending on the sign of
$a$, $b$, etc., the $R_a$, $R_b$, etc. are either the R-charge
or two minus the R-charge.  For example, for the $Ai$
quiver
$R(X_{21}) = R_a$, $R(X_{32})= R_d$, 
$R(X_{43}) = R_f$, and $R(X_{42}) = R_e$.
However, 
$R(X_{13}) = 2- R_b$ and $R(X_{14}) = 2- R_c$.

We expect the R-charges to be positive, and it is interesting
to investigate whether the constraints so far
considered enforce this positivity.
Indeed from the sign assignments (\ref{signassign}) it
is clear that for the $Ai$ quiver, 
$R(X_{21})$, $R(X_{32})$, and $R(X_{43})$ are all positive.
It is straightforward, using (\ref{signassign}) and the 
rank formulae (\ref{rels}) to show that the remaining R-charges
are also positive.

For the $F$ type quivers, the R-charges may in general be negative, as we will
now see.  Take
the $Fii$ quiver:
$R(X_{21}) = R_a$, $R(X_{32})= R_d$,  $R(X_{43}) = R_f$,
and $R(X_{41}) = R_c$ while
$R(X_{13}) = 2-R_b$ and $R(X_{24}) = 2-R_e$. 
While it is straightforward to show that $R(X_{21})$, $R(X_{32})$, $R(X_{43})$, and
$R(X_{41})$ are positive, the remaining two R-charges may in general be
negative.  Take for example the $Fii$ type quiver
\be
S = \left( 
\begin{array}{rrrr}
1 & -2 & 5 & -3 \\
0 & 1 & -7 & 5 \\
0 & 0 & 1 & -2 \\
0 & 0 & 0 & 1
\end{array}
\right)
\ee
with gauge group ranks all $SU(N)$.
This ``gauge theory'' has
has $R(X_{13})=R(X_{24})=-1/4$ according to our formulae.
Gauge invariant, antisymmetric products of these 
$X_{13}$ and $X_{24}$ correspond to dibaryonic
operators and naively would have a negative R-charge.
As these dibaryons are chiral primaries, they would also
have a negative conformal dimension 
in violation of unitarity.\footnote{
K.~Intriligator and R.~Tatar have suggested in private communication
that 
these negative R-charges
might indicate the appearance of an accidental U(1) symmetry in 
the gauge theory.
In simpler gauge theories
where this problem occurs, for example SQCD,
as a result of the symmetry the troublesome gauge invariant 
operators decouple from the interacting theory and become free chiral 
superfields. 
(See \cite{Kutasov} for a recent discussion of this phenomenon.)
However, it is not clear to the author what would become of these troublesome dibaryons.}

\subsection{Mutation versus Seiberg Duality}

We investigate the effects of mutation and Seiberg Duality on a four node quiver.  Up to now, we have considered $L^D$ and $R^D$ type mutations.
To mutate a collection to obtain a different collection, 
we operate with $R$ and $L$ type mutations on the helix ${\mathcal H}$.
In this way, we can be sure that the ranks of the gauge groups
stay positive.  If we were to mutate the dual gauge theory collection, 
where some
of the sheaves have negative rank, it is not {\it a priori} 
clear how to choose the
signs to satisfy chiral anomaly cancellation.

For the four node quivers, then, the
geometric collection $\cale^G = (E_4^\vee, E_3^\vee, E_2^\vee, E_1^\vee)$
generates the helix $\calh$.  Consider $L_{E^\vee_4} E_3^\vee$.
Under such a mutation, the entries of $S$ become
\begin{eqnarray}
a \to a \ , &  b \to c-bf \ , & c \to \pm b \ , \nonumber \\
d \to e-df \ , & e \to \pm d \ , & f \to \mp f \ .
\label{leftmfn}
\end{eqnarray}
One may also consider $R_{E_3^\vee} E_4^\vee$ where
\begin{eqnarray}
a \to a \ , &  b \to \pm c \ , & c \to b-cf \ , \nonumber \\ 
d \to \pm e \ , & e \to d-ef \ , & f \to \mp f \ .
\label{rightmfn}
\end{eqnarray}
Applying these transformations to $Ai$, $Fi$, $Aii$, etc.,
one finds that mutations in general map $A$ type quivers to
both $A$ and $F$ type quivers and similarly for $F$ type.

However, it is worthwhile to look more closely.
In \cite{unify}, it was pointed out that
Seiberg duality sometimes corresponds to a mutation.
Seiberg duality for us will mean a particular combinatoric
action on the quiver.  A careful treatment requires knowledge
of the superpotential which we lack in general.  
Combinatorially, 
we specify a node on which to dualize.  We change the rank
of the gauge group at that node from $N_c$ to $N_f-N_c$.
For example, for an $Ai$ quiver, dualizing on node 4
would send $N r_4 \to N(c r_1 - r_4)$.   
For every bifundamental that transformed under $SU(N_c)$, 
we introduce
a new bifundamental with the opposite chirality that transforms under
$SU(N_f-N_c)$.    
For $Ai$ and node 4, this introduction would send $c\to -c$,
$e\to -e$, and $f \to -f$.  Finally, we 
combine the old bifundamentals that transformed under
$SU(N_c)$ into mesonic type operators.
These mesonic operators
look like new bifundamentals
and ensure chiral anomaly cancellation.
The superpotential is critical at this step;
unless the superpotential allows these bifundamentals
to be integrated out properly, bidirectional arrows
may exist in the Seiberg dual theory, spoiling a
description using exceptional collections.
Assuming the appropriate superpotential, for $Ai$ and node 4, 
these mesonic operators send $b \to b-cf$ and $a \to a-ce$.

Now Seiberg duality does not respect the ordering of the
quiver.  After a Seiberg duality, the quiver must usually be reordered
in order to correspond to an exceptional collection.

Returning to our investigation of the relation between
Seiberg duality and mutation,
recall that in general each node in a four node quiver
is connected to the other three nodes by
three arrows; not all the arrows point in the same direction.

Assume only one arrow points into the node.  If this arrow comes from a node that is a nearest neighbor to the right
in the gauge theory collection $\cale^Q$, then
a left
mutation over the corresponding node in the geometric collection
will correspond to
a Seiberg duality. 

Assume only one arrow points away from the node.  If this arrow goes to
a node that is a nearest neighbor to the left in the gauge theory collection 
$\cale^Q$, then
a right mutation over this neighboring node in the geometric collection 
corresponds to Seiberg
duality.

Moreover, the mutation whether left or right will always be a division in these
special circumstances. 

In general, there are eight possible mutations, left and right for each of the four nodes.  However, there are only four ordinary Seiberg dualities, one for
each node.  Looking closely at the $A$ and $F$ type quivers we see that
for $A$ type, four of the eight mutations correspond to Seiberg dualities.
In other words, Seiberg duality can always be thought of as a mutation
for the $A$ type quiver.  Specifically, for the $A$ type quiver, every node
has a single in or out arrow pointing to or away from a neighboring node in the 
collection.  Moreover, the single in arrow will come from the right and the single
out arrow will always point to the left in $\cale^Q$.

For the $F$ type quiver, only two of the
eight mutations correspond to Seiberg dualities.  The remaining two
Seiberg dualities don't correspond to mutations.  More precisely,
assuming that no arrows become bidirectional after the duality,
we cannot satisfy the constraint (\ref{I}) for any ordering after the
duality.
It seems likely that if these $F$ type quivers have a sensible
gauge theory interpretation, then 
these remaining two Seiberg dualities produce
quivers with bidirectional arrows, i.e.~quivers that cannot possibly
come from exceptional collections.

One remaining intriguing fact here is that Seiberg duality, when
it corresponds to a mutation, maps $A$ type quivers to $A$ type 
quivers and $F$ type to $F$ type.

%

Based on the negative R-charges and the strange behavior under Seiberg duality, it is 
very tempting to conclude that $F$ type quivers are not allowed.  However, we lack
a geometric understanding of why these $F$ type quivers should be ruled out.
From the exceptional collection point of view, one geometric collection seems
just as good as another.\footnote{
To the author's knowledge, the first example of such an $F$ type quiver appeared in
\cite{PicLef}.}

As mentioned in the introduction, it is tempting to generalize from our experience
with four node quivers.  We define a {\it well split} quiver to be such that
for any node $i$, all the nodes in-going into $i$ can be placed to the right in the
$\cale^Q$  and all the outgoing nodes with respect to $i$
to the left in $\cale^Q$.  For such a quiver, Seiberg duality should always
correspond to a left mutation of $E_i^G$ over all the in-going nodes 
in $\cale^G$ \cite{wijn, HW}.  Furthermore, one hopes that the Seiberg
dual of a well split quiver is again well split.
In this language, $A$ type four-node quivers are well split.  
In constrast, a quiver which does not satisfy this property we call {\it ill split}
because the in-going and outgoing nodes do not split in a way that
respects the ordering of the exceptional collection.
$F$ type quivers are ill split quivers.

\section{Toward a General Understanding}

In proving cyclic invariance of the quiver ordering, deriving the formula for
the ranks of the gauge groups in terms of paths in the quiver,  
classifying four node quivers, and checking
the value of $\Tr R^3$ for four node quivers, 
we have taken a few steps toward
understanding the connections between exceptional collections,
del Pezzo gauge theories, and AdS/CFT correspondence.
However, there remains a long list of things to do.
In hopes of inspiring the reader, we briefly describe
a few of these items.

Number one on this list is obtaining a better understanding of
$F$ type quivers, or more generally, ill split quivers.  For four node
quivers, we saw that the set of $A$ type quivers was closed
under Seiberg duality.  
Thus, we can consider only $A$ type quivers if we are
interested in their behavior under RG flow.  More generally,
it would be interesting to prove that well split quivers
are closed under Seiberg duality.  Also interesting would
be some kind of geometric demonstration that 
$F$ type and ill split
quivers are not allowed as gauge theories.

Number two is an analysis of the behavior
of these gauge theories under RG flow.
How generic is the KS flow of \cite{klst}? 
Can we obtain an analytic understanding of the 
duality walls in \cite{fhhk}?  We plan to return
to these questions in \cite{FHHW}. 

Number three  
would be to understand how
exceptional collections relate to
another way of thinking of
Seiberg duality considered in
\cite{BD,Braun,indians}.  
In these papers, Seiberg duality is related
to tilting equivalences of certain
derived categories.

Finally, there are a large number of technical details which
need to be resolved.  
Here are three such details.  
One ought to
be able to prove that the R-charge formula (\ref{Req})
corresponds to the maximization of
$a_c$ principle derived by \cite{IW}.  Second, one ought
to be able to
use the R-charges to check 
that $\Tr R^3 = 24/K^2$ for a general del Pezzo quiver.
Third, one should show that the rank
two condition and the trace condition $\Tr S S^{-T}=n$ on
$S$ are not only necessary but sufficient for 
$S$ to correspond to an exceptional collection.

Hopefully, some of these issues will be resolved soon.

\section*{Acknowledgments} 
It is a pleasure to thank S.~Franco, 
A.~Hanany, 
Y.-H.~He, 
K.~Intriligator, 
J.~M\raise 3pt \hbox{\text {\normalsize c}}Kernan, 
M.~Rangamani,
M.~Spradlin, and 
R.~Tatar
for discussion.  The author would
like to thank J.~Walcher for comments on the
manuscript.  The author would also
like to thank Berkeley, where part of
this work was prepared, for hospitality.
This
research was supported in part by the National Science Foundation under
Grant No.~PHY99-07949.


\begin{thebibliography}{99}

\bi{MoPle}
D.~Morrison and R.~Plesser,
``Non-Spherical Horizons, I,''
{\it Adv. Theor. Math. Phys.} {\bf 3} (1999) 1,  {\tt hep-th/9810201}.

\bi{acharya}
B.~S.~Acharya, J.~M.~Figueroa-O'Farrill, C.~M.~Hull,
and B.~Spence,
``Branes at conical singularities and holography,''
{\it Adv. Theor. Math. Phys.} {\bf 2} (1999) 1249,
{\tt hep-th/9808014}.

\bibitem{KW}
I.~R.~Klebanov and E.~Witten, ``Superconformal field theory on three-branes
at a Calabi-Yau singularity,'' 
{\it Nucl. Phys.} {\bf B536} (1998) 199, 
{\tt hep-th/9807080}.

\bi{Kehag}
A. Kehagias, ``New Type IIB Vacua and Their F-Theory Interpretation,''
{\it Phys. Lett.}  {\bf B435} (1998) 337, 
{{\tt hep-th/9805131}}.

\bibitem{jthroat}
J.~Maldacena, ``The Large N limit of superconformal field theories and
supergravity,'' {\it Adv. Theor. Math. Phys.} {\bf 2} (1998) 231, 
{{\tt hep-th/9711200}}.

\bibitem{gkp}
S.S. Gubser, I.R. Klebanov, and A.M. Polyakov, ``Gauge theory correlators
from noncritical string theory,''
{\it Phys. Lett.} {\bf B428} (1998) 105,
{{\tt hep-th/9802109}}.

\bibitem{EW}
E.~Witten, ``Anti-de Sitter space and holography,''
{\it Adv. Theor. Math. Phys.} {\bf 2} (1998) 253,
{{\tt hep-th/9802150}}.

\bibitem{beasley0}
C.~Beasley, B.~R.~Greene, C.~I.~Lazaroiu and M.~R.~Plesser,
``D3-branes on partial resolutions of abelian quotient singularities of  Calabi-Yau threefolds,''
{\it Nucl. Phys.} {\bf B566} (2000) 599,
{\tt hep-th/9907186}.

\bi{Beasley1}
C.~E.~Beasley and M.~Ronen~Plesser, 
``Toric Duality is Seiberg Duality,''
{\it JHEP} {\bf 0112} (2001) 001,
{\tt hep-th/0109053}

\bibitem{fhh1}
B.~Feng, A.~Hanany and Y.~H.~He,
``D-brane gauge theories from toric singularities and toric duality,''
{\it Nucl. Phys.} {\bf B595} (2001) 165,
{\it hep-th/0003085}.

\bibitem{fhh2}
B.~Feng, A.~Hanany and Y.~H.~He,
``Phase structure of D-brane gauge theories and toric duality,''
{\it JHEP} {\bf 0108} (2001) 040,
{\tt hep-th/0104259}.

\bibitem{unify}
F.~Cachazo, B.~Fiol, K.~A.~Intriligator, S.~Katz and C.~Vafa,
``A geometric unification of dualities,''
{\it Nucl. Phys.} {\bf B 628} (2002) 3,
{\tt hep-th/0110028}.

\bibitem{wijn}
M.~Wijnholt,
``Large volume perspective on branes at singularities,''
{\tt hep-th/0212021}.

\bi{ffhh}
B.~Feng, S.~Franco, A.~Hanany, and Y.~H.~He, 
``Unhiggsing the del Pezzo,'' 
{\it JHEP} {\bf 0308} (2003) 058,
{\tt hep-th/0209228}.

\bibitem{haiq}
A.~Iqbal and A.~Hanany, 
``Quiver Theories from D6-branes via Mirror Symmetry,''
{\it JHEP} {\bf 0204} (2002) 009,
{\tt hep-th/0108137}

\bibitem{HW}
C.~P.~Herzog and J.~Walcher, ``Dibaryons from Exceptional Collections,''
{\it JHEP} {\bf 0309} (2003) 060,
{\tt hep-th/0306298}.

\bi{HM}
C.~P.~Herzog and J.~M\raise 3pt \hbox{\text {\normalsize c}}Kernan,
``Dibaryon Spectroscopy,''
{\it JHEP} {\bf 0308} (2003) 054,
{\tt hep-th/0305048}.

\bibitem{IW2}
K.~Intriligator and B.~Wecht,
``Baryon charges in 4D superconformal field theories and their AdS  duals,''
{\tt hep-th/0305046}.



\bibitem{klst}
I.~R.~Klebanov and M.~J.~Strassler,
``Supergravity and a confining gauge theory: Duality cascades and  chiSB-resolution of naked singularities,''
JHEP {\bf 0008}, 052 (2000) 052,
{\tt hep-th/0007191}.

\bibitem{FHHW}
S.~Franco, Y.-H.~He, C.~P.~Herzog, J.~Walcher, 
``Chaotic Duality in String Theory,'' {\tt hep-th/0402120}.

\bibitem{rudakov}
``Helices and vector bundles,''
Seminaire Rudakov. London Mathematical Society Lecture Note Series, 148. 
Cambridge University Press, Cambridge, 1990.

\bibitem{NK}
B. V. Karpov\ and\ D. Yu. Nogin, 
``Three-block Exceptional Collections over del Pezzo Surfaces,''
Izv. Ross. Akad. Nauk Ser. Mat. {\bf 62} (1998), no. 3, 3--38; 
translation in Izv. Math. {\bf 62} (1998), no.~3, 429--463

\bibitem{hiv}
K.~Hori, A.~Iqbal and C.~Vafa,
``D-branes and mirror symmetry,''
{\tt hep-th/0005247}.

\bibitem{KO}
S. A. Kuleshov\ and\ D. O. Orlov, 
``Exceptional sheaves over del Pezzo surfaces,''
Izv. Ross. Akad. Nauk Ser. Mat. {\bf 58} (1994), no. 3, 53--87; 
translation in Russian Acad. Sci. Izv. Math. {\bf 44} (1995), no.~3, 479--513 

\bibitem{anselmi1}
D.~Anselmi, D.~Z.~Freedman, M.~T.~Grisaru and A.~A.~Johansen,
``Nonperturbative formulas for central functions of supersymmetric gauge  theories,''
{\it Nucl. Phys.} {\bf B526} (1998) 543,
{\tt hep-th/9708042}.

\bibitem{anselmi2}
D.~Anselmi, J.~Erlich, D.~Z.~Freedman and A.~A.~Johansen,
``Positivity constraints on anomalies in supersymmetric gauge theories,''
{\it Phys. Rev. D} {\bf 57} (1998) 7570,
{\tt hep-th/9711035}.

\bibitem{skenderis} {M.~Henningson and K.~Skenderis,
``The holographic Weyl anomaly,''
{\it JHEP} {\bf 9807} (1998) 023,
{\tt hep-th/9806087}.}

\bibitem{gubser}
S.~S.~Gubser, ``Einstein Manifolds and Conformal Field Theories,'' 
{\it Phys. Rev. D} {\bf 59} (1999) 025006, 
{\tt hep-th/9807164}.

\bibitem{BH}
A.~Bergman and C.~P.~Herzog,
``The volume of some non-spherical horizons and the AdS/CFT
correspondence,''
{\it JHEP} {\bf 0201} (2002) 030,
{\tt hep-th/0108020}.

\bibitem{Kutasov}
D.~Kutasov, A.~Parnachev, and D.~A.~Sahakyan,
``Central Charges and $U(1)_R$ Symmetries
in ${\mathcal N}=1$ Super Yang-Mills,''
{\tt hep-th/0308071}.

\bibitem{PicLef}
B.~Feng, A.~Hanany, Y.~H.~He and A.~Iqbal,
``Quiver theories, soliton spectra and Picard-Lefschetz transformations,''
{\it JHEP} {\bf 0302} (2003) 056,
{\tt hep-th/0206152}.

\bibitem{fhhk}
S.~Franco, A.~Hanany, Y.-H.~He, and P.~Kazakopoulos,
``Duality Walls, Duality Trees and Fractional Branes,''
{\tt hep-th/0306092}.

\bibitem{BD}
D.~Berenstein and M.~R.~Douglas,
``Seiberg duality for quiver gauge theories,''
{\tt hep-th/0207027}.

\bibitem{Braun}
V.~Braun,
``On Berenstein-Douglas-Seiberg duality,''
{\it JHEP} {\bf 0301} (2003) 082,
{\tt hep-th/0211173}.

\bibitem{indians}
S.~Mukhopadhyay and K.~Ray,
``Seiberg duality as derived equivalence for some quiver gauge theories,''
{\tt hep-th/0309191}.

\bibitem{IW}
K.~Intriligator and B.~Wecht,
``The exact superconformal R-symmetry maximizes a,''
{\it Nucl. Phys.} {\bf B667} (2003) 183,
{\tt hep-th/0304128}.

\end{thebibliography}
\end{document}